\documentclass[fleqn,usenatbib,onecolumn]{mnras}
\usepackage{mathptmx,times}
\usepackage[T1]{fontenc}
\DeclareRobustCommand{\VAN}[3]{#2}
\let\VANthebibliography\thebibliography
\def\thebibliography{\DeclareRobustCommand{\VAN}[3]{##3}\VANthebibliography}
\usepackage{graphicx}	
\usepackage{amsmath}	

\newcommand{\rone}{FRB~121102}
\newcommand{\rthree}{FRB~180916}

\newcommand{\sgra}{Sgr~A*}
\newcommand{\gcmag}{J1745$-$2900}
\newcommand{\sgr}{SGR~1935+2154}

\newcommand{\RM}{\rm RM}
\newcommand{\DM}{\rm DM}
\newcommand{\DMunits}{{\rm pc~cm^{-3}}}

\title[Fast Radio Bursts]{Progress in Understanding the Enigmatic Fast Radio Bursts}
\author[S. Chatterjee]{Shami Chatterjee$^{1}$\thanks{E-mail: shami@astro.cornell.edu} \\
$^{1}$Cornell Center for Astrophysics and Planetary Science, Cornell University, Ithaca, NY 14853, USA
}
\date{Invited review for {\em Astronomy \& Geophysics}}
\pubyear{2020}

\begin{document}
\label{firstpage}
\pagerange{\pageref{firstpage}--\pageref{lastpage}}
\maketitle

\begin{abstract}

In less than a decade, fast radio bursts have gone from a single debated curiosity to a diverse extragalactic population with established host galaxies and energy scales. While a wide range of models remain viable, the central engines of FRBs are likely to involve energetic young magnetars, as confirmed by the recent discovery of a Galactic analog to these extragalactic bursts. Here we provide a brief introductory review of fast radio bursts, focusing on the rapid recent progress in observations of these enigmatic events, our understanding of their central engines, and their use as probes of the intergalactic medium. We caution against a rush to judgement on the mechanisms and classification of all FRBs: at this point, it remains plausible that there could be one dominant central engine, as well as the possibility that radio bursts are a generic feature produced by many different mechanisms. We also emphasize the importance of improved modeling of our Galaxy and Galactic halo, which otherwise impose systematic errors on every FRB line of sight. The future of science with fast radio bursts appears bright.

\end{abstract}




\section{Introduction and Overview}

The time-domain radio sky is no longer {\em terra incognita}. Spurred on by advances in radio telescope sensitivity, bandwidth, resolution, and survey cadence, as well as huge leaps forward in computational capability and data storage, we now know of time variable phenomena on timescales spanning from years down to nanoseconds \citep{hkwe03}, and operating on physical scales from parsecs to meters.

While periodic short-duration radio pulses have been known since the discovery of radio pulsars \citep{hbp+68}, searches for isolated single pulses remained inconclusive \citep[e.g.,][]{1980ApJ...236L.109L,alv89} until the discovery of rotating radio transients \citep[RRATs,][]{mll+06}, which are now understood as Galactic pulsars with only occasionally-detectable pulses. These pulses all show a consistent dispersion, arriving earlier at high frequencies and later at lower frequencies due to their propagation through the cold tenuous plasma of interstellar space (see \S\ref{sec:dispers} and Figure~\ref{fig:dispers} below), with an amount of dispersion consistent with an origin within our Milky Way galaxy.

Fast radio bursts are dispersed, isolated, millisecond-duration radio pulses similar in appearance to single pulses from Galactic pulsars, with the defining characteristic of a pulse dispersion measure that exceeds the maximum expected from our Galaxy in the originating direction. The first discovered fast radio burst (FRB) was identified in archival Parkes Multibeam survey data by \citet[][see Figure~\ref{fig:ds} below]{lbm+07}, but the extragalactic origin suggested by its exceptional degree of pulse dispersion implied an enormous intrinsic brightness, making its astrophysical nature controversial, at least until further examples of FRBs were identified by \citet{tsb+13}.

Given the limited instantaneous field of view and on-sky time of typical high-time-resolution radio surveys, the detection of even a small number of short duration transient events is unlikely, and suggests that the true all-sky rate is enormous. \citet{tsb+13} estimated a (detection threshold-dependent) event rate $\sim 1^{+0.6}_{-0.5} \times 10^4$~sky$^{-1}$~day$^{-1}$, leading to a veritable gold rush of theoretical modeling efforts for central engines that accommodated both the enormous event rates and the high energy requirements for these millisecond flashes to be visible at cosmological (or at least extragalactic) distances.

The discovery of a repeating FRB at the Arecibo Observatory \citep[FRB~20121102A;][]{sch+14,ssh+16a} immediately argued against a cataclysmic central engine, at least for some subset of FRBs, while its localization at the Karl G. Jansky Very Large Array \citep{clw+17} and the measurement of the redshift of its dwarf host galaxy, $z\sim0.2$, \citep{tbc+17} confirmed the extragalactic nature of these bursts and allowed a firm estimate of the isotropic equivalent energy budget for FRBs of $\sim$10$^{38}$~erg. Since then, observations have progressed rapidly, with a large number of extremely bright bursts identified at the Australian SKA Pathfinder (ASKAP) telescope \citep{smb+18}, further localizations and host galaxy determinations \citep[e.g.,][]{bdp+19,rcd+19}, and the Canadian Hydrogen Intensity Mapping Experiment (CHIME) detecting a bonanza of events \citep[e.g.,][]{chime19frbs}, including further repeating sources \citep{chime19repeater, chime19reps}. Meanwhile, although early theoretical models spanned the gamut from local flare stars \citep{lsm14, mls+15a} to cosmic strings \citep{vach08}, the most well-developed models feature very young neutron stars with extreme magnetic fields (magnetars, $B>10^{14}$~G) embedded in their birth supernova remnants and/or their wind nebulae, with FRB emission arising from relativistic shocks in these nebulae \citep[e.g.,][]{lyubarsky14,mm18,mms20,beloborodov20}.

Here we provide a brief and selective introduction to fast radio bursts, building on comprehensive recent reviews by \citet{cc19} and \citet{phl19}. \citet{z20} provides a recent compact overview of theoretical models, while an extensive collection of models is curated by \citet{pww+19}. For a blow-by-blow view from the front lines, we refer interested readers to the FRB Community Newsletter\footnote{FRB Community Newsletter: \url{http://frb.astro.cornell.edu/news/}}. Below we summarize propagation effects relevant to understanding FRBs (\S2) and provide an overview of the FRB population (\S3), repeating FRBs (\S4), and host galaxy identifications and the use of FRBs as cosmological probes (\S5). Finally, we describe a Galactic FRB event and its implications for the FRB central engine (\S6), and conclude with an overview of future prospects for the field (\S7).

\section{Propagation Effects}
\label{sec:dispers}

\begin{figure}
	\includegraphics[width=0.96\textwidth]{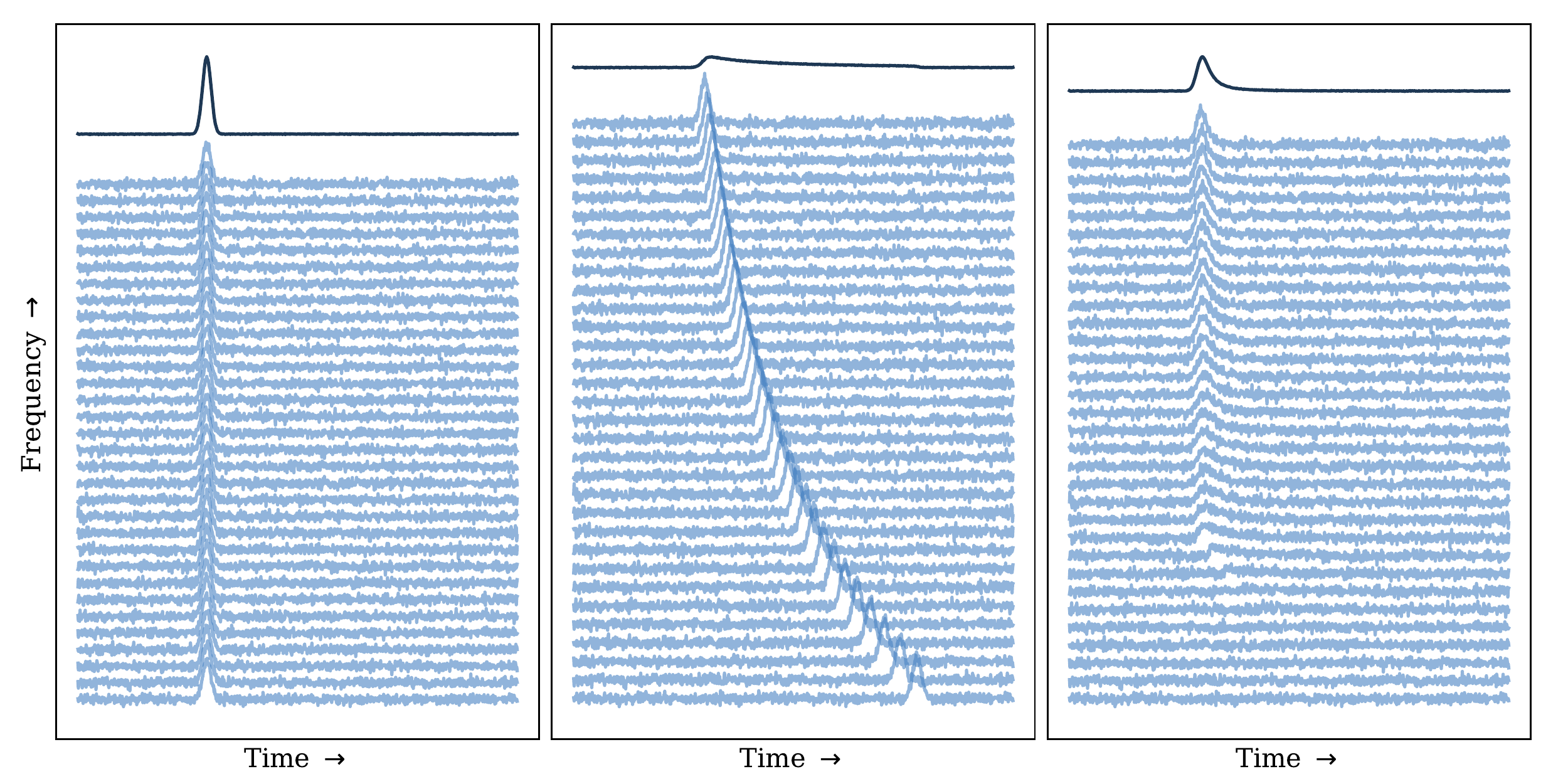}
    \caption{A burst or pulse (left panel) is dispersed (middle panel) and scattered (right panel) by propagation through an ionized medium. While both effects are frequency-dependent, dispersion is deterministic and can be reversed, while scattering can be understood as a convolution of the pulse with an interstellar medium transfer function, which does not necessarily allow for a unique deconvolution.}
    \label{fig:dispers}
\end{figure}

{\bf Pulse dispersion:} The interstellar medium (ISM) and intergalactic medium (IGM) both consist of cold, tenuous, magnetized plasma, which imposes a frequency-dependent delay on the time of arrival of a pulse, $\Delta t \propto \DM \,\nu^{-2}$, where the dispersion measure \DM\ is the integrated column density of electrons $n_e$ along the line of sight:
\[
\DM = \int_0^L n_e \, ds,
\]
weighted by $(1+z)^{-1}$ for the cosmological case,                                                                                                                                                                            . As illustrated in Figure~\ref{fig:dispers}, dispersion results in a smeared-out pulse in the time domain, making detection unlikely unless the \DM\ is fit for and removed first. If the baseband radio signal can be sampled with high enough time resolution, the effect of a known (or trial) DM can be deterministically removed. Otherwise, the channelized time-frequency data can be searched over in increments of DM, at the cost of some residual intra-channel smearing that reduces the significance of a detected pulse. The latter is more typical for present-day blind searches, due to the computational cost of working with baseband data.

Pulses from Galactic radio pulsars show consistent values of DM over long periods of time (with small variations due to interstellar ``weather'') and with enough pulsars at known distances, the known values of the integrated electron column density can be used to map out the Galactic electron density distribution \citep{tc93}. Electron density distribution models such as NE2001 \citep{cl02} and YMW16 \citep{ymw17} are widely used to determine whether the DM of a pulse or burst can be accounted for within the Milky Way, or if it is more likely to be of extragalactic origin. After accounting for the contributions of the Milky Way disk and halo, as well as the ISM of the host galaxy, extragalactic burst DMs quantify the integrated column density of the IGM, allowing for a census of the baryons in that tenuous gas \citep[e.g.,][]{mpm+20}.

{\bf Scattering and Scintillation:} Electron density fluctuations on scales larger than a wavelength can lead to small-angle scattering, leading to strongly frequency-dependent pulse scattering ($\propto \nu^{-4}$), as shown in Figure~\ref{fig:dispers}. That can be understood as multi-path propagation leading to a smearing of the pulse (as well as angular broadening of the source image), or as a convolution of the pulse or burst with a (frequency-dependent) propagation transfer function in the time domain. Like other convolution processes, scattering does not necessarily allow for unique deconvolution.

Along with scattering, pulses can show both refractive and diffractive scintillation due to propagation through a turbulent medium. Diffractive interstellar scintillation, the radio analog of stellar twinkling due to turbulence in the Earth's atmosphere, can produce up to 100\% modulation in the spectrum of a burst, producing band-limited structures in the dynamic spectrum. Scintillation and scattering effects are often apparent in the spectra of FRBs, as shown in Figure~\ref{fig:ds}, and have been used to probe the turbulence spectrum of the ionized IGM \citep[e.g.,][]{rsb+16}.

{\bf Plasma Lensing:} Density structure in the host galaxies of FRBs can act as plasma lenses, producing frequency-dependent focusing and defocusing at the detector, as described by \citet{cwh+17}. Such lensing would be the extragalactic equivalent of the Galactic ``extreme scattering events'' \citep{fdjh87}.
Large magnification factors up to $\sim10^2$ are possible over short timescales (hours to days) and narrow frequency ranges (0.1--1 GHz), if the detector location falls on a caustic.

{\bf Faraday Rotation:} The left- and right-hand circularly polarized components of a radio wave have different refractive indices in the magnetized plasma of the ISM and IGM, and travel at different speeds, leading to a wavelength-dependent rotation of the observed linear polarization angle $\theta$ compared to the angle at emission, $\Delta \theta = \RM \, \lambda^2$, for wavelength $\lambda$. RM, the rotation measure, is given by the integral of the component of the magnetic field along the line of sight, weighted by the electron density:
\[
\RM \propto \int_0^L B_{||} n_e \,ds.
\]
Thus a burst or pulse with detectable linear polarization can be used to trace the integrated density-weighted magnetic field along the line of sight, including the magnetic field in the IGM and the host galaxy environment \citep{mls+15b,rsb+16,msh+18}.

\section{The Population of Fast Radio Bursts}

\begin{figure}
	\includegraphics[width=0.3\textwidth]{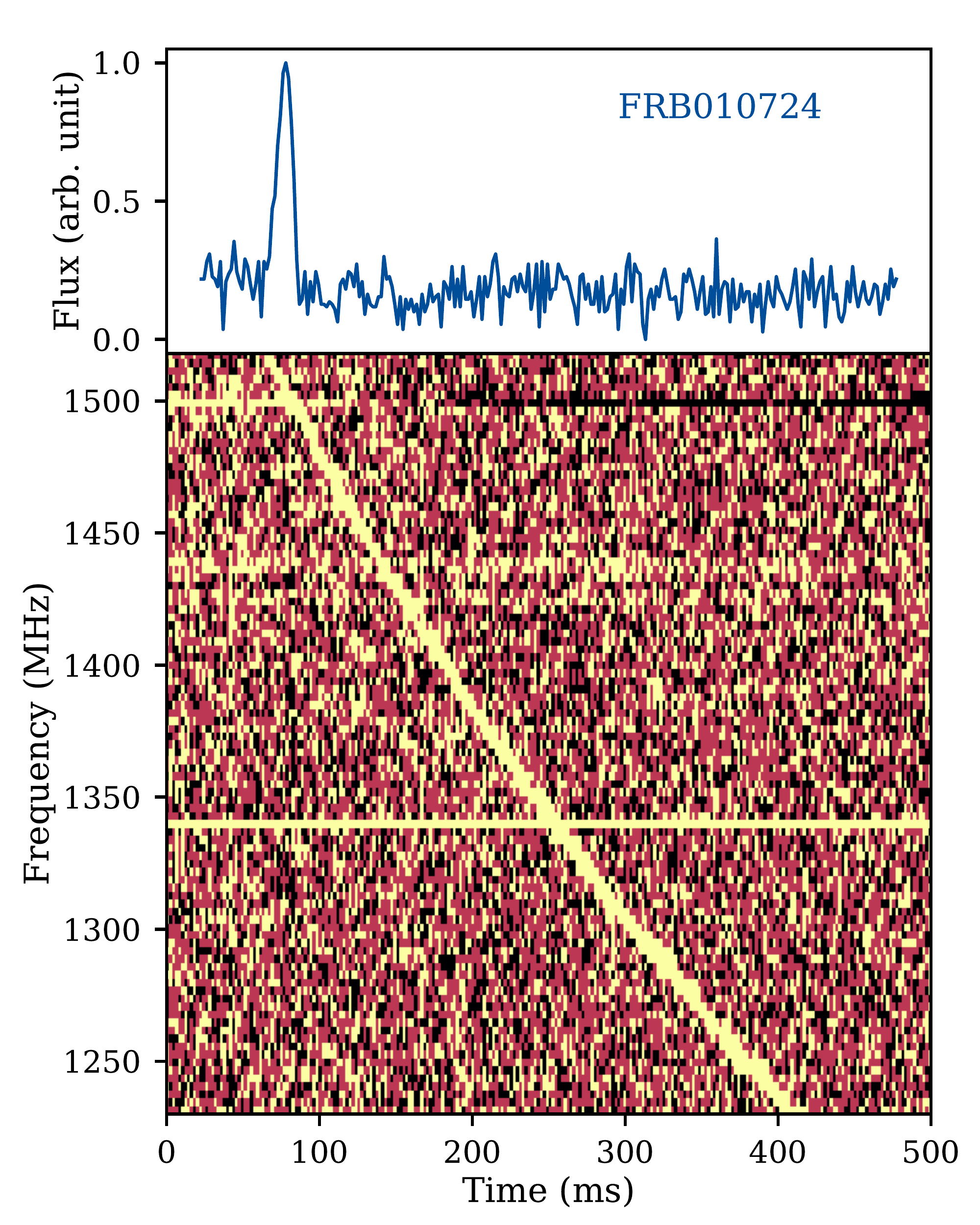}
	\includegraphics[width=0.3\textwidth]{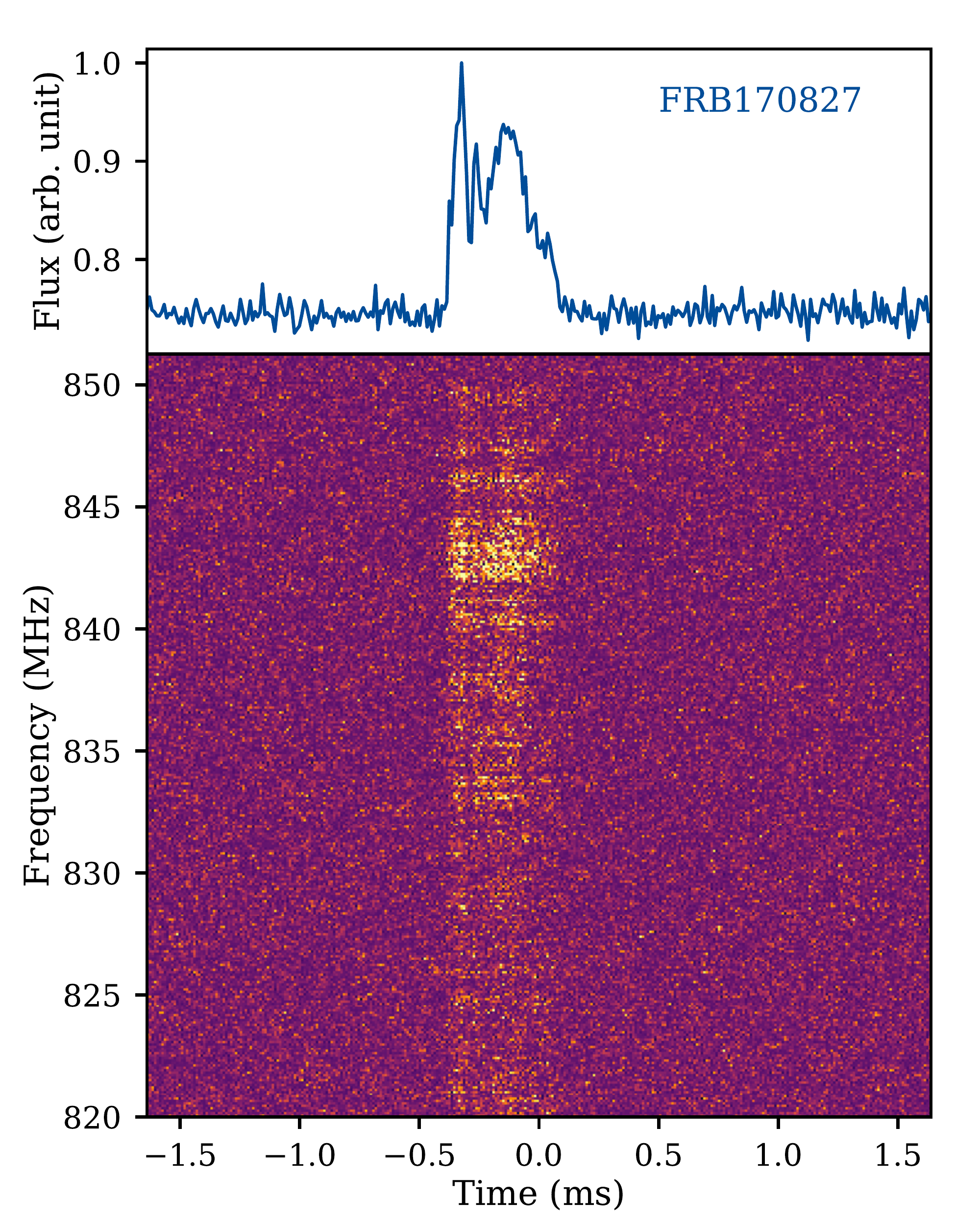}
	\includegraphics[width=0.3\textwidth]{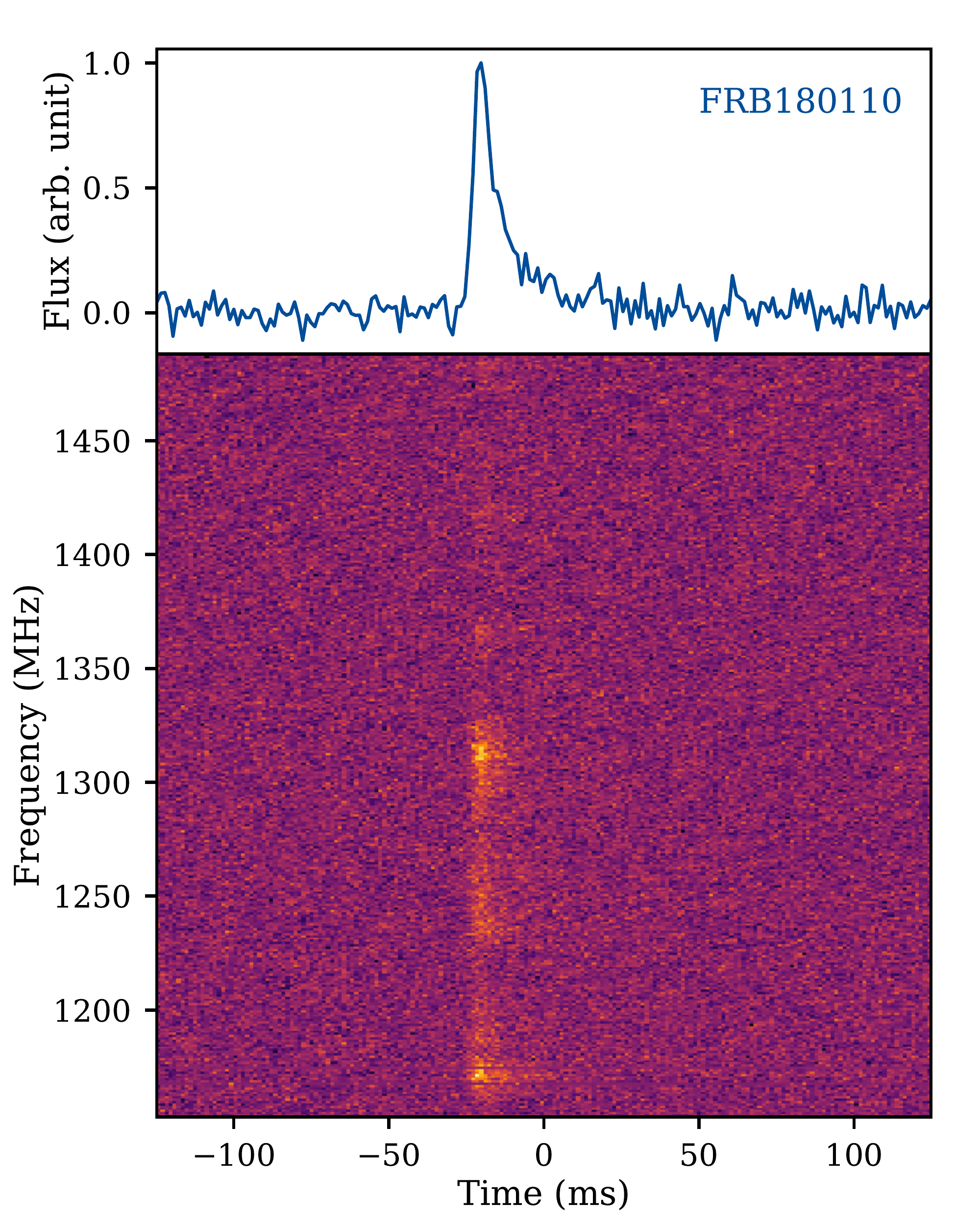}
    \caption{Examples of fast radio burst dynamic spectra.
    Left: FRB~010724, the first reported example \citep{lbm+07}, with DM $=375~\DMunits$. The lower panel shows the burst dispersion sweep in the time-frequency plane; after removing the best fit dispersion sweep, the frequency-averaged burst profile is shown in the upper panel.
    Middle: FRB~170827, detected at UTMOST \citep{ffb+18}. The lower panel shows the dynamic spectrum after removal of the dispersion sweep, DM $=177~\DMunits$. Voltage capture was triggered after real-time detection, revealing time-frequency scintillation structure in the burst after coherent de-dispersion. The upper panel shows the frequency-averaged burst profile.
    Right: FRB~180110, a bright burst detected at ASKAP \citep{smb+18} and shown after removal of the dispersion sweep with DM $=716~\DMunits$. Note the scattering tail and the scintillation structure.}
    \label{fig:ds}
\end{figure}

Standard searches for single pulses and bursts operate on dynamic spectra acquired at radio telescopes with high time and frequency resolution. The data are typically cleaned of interference, calibrated for the instrumental bandpass response, de-dispersed at trial values of DM, and frequency averaged, with the resulting time series being inspected for departures from random noise. Advances in computational capability and storage capacity have allowed for the exploration of many variations on this basic approach, including the development of the fast discrete dispersion measure transform \citep[FDMT;][]{zo17} and the use of Graphical Processing Units, machine learning, and computer vision approaches \citep[e.g., {\sc Fetch};][]{aab+20}. Real-time searches have allowed for the capture of buffered voltage data, allowing for coherent dedispersion of newly-detected FRBs (e.g., \citealt{pfg+19,ffb+18} and Figure~\ref{fig:ds}).

Radio frequency interference (RFI) is the bane of all of these searches. With interference sources ranging from the local (unshielded electronics and computers, automobile spark plugs) to the global (constellations of low-earth orbit satellites), the radio spectrum is an increasingly endangered resource overwhelmed by commercial pressure. Beyond relying on the dispersive sweep of astrophysical signals, RFI excision algorithms have become more sophisticated over time, but transient signals are frequently swamped by false positive candidates from interference.

\begin{figure}
	\includegraphics[width=0.48\textwidth]{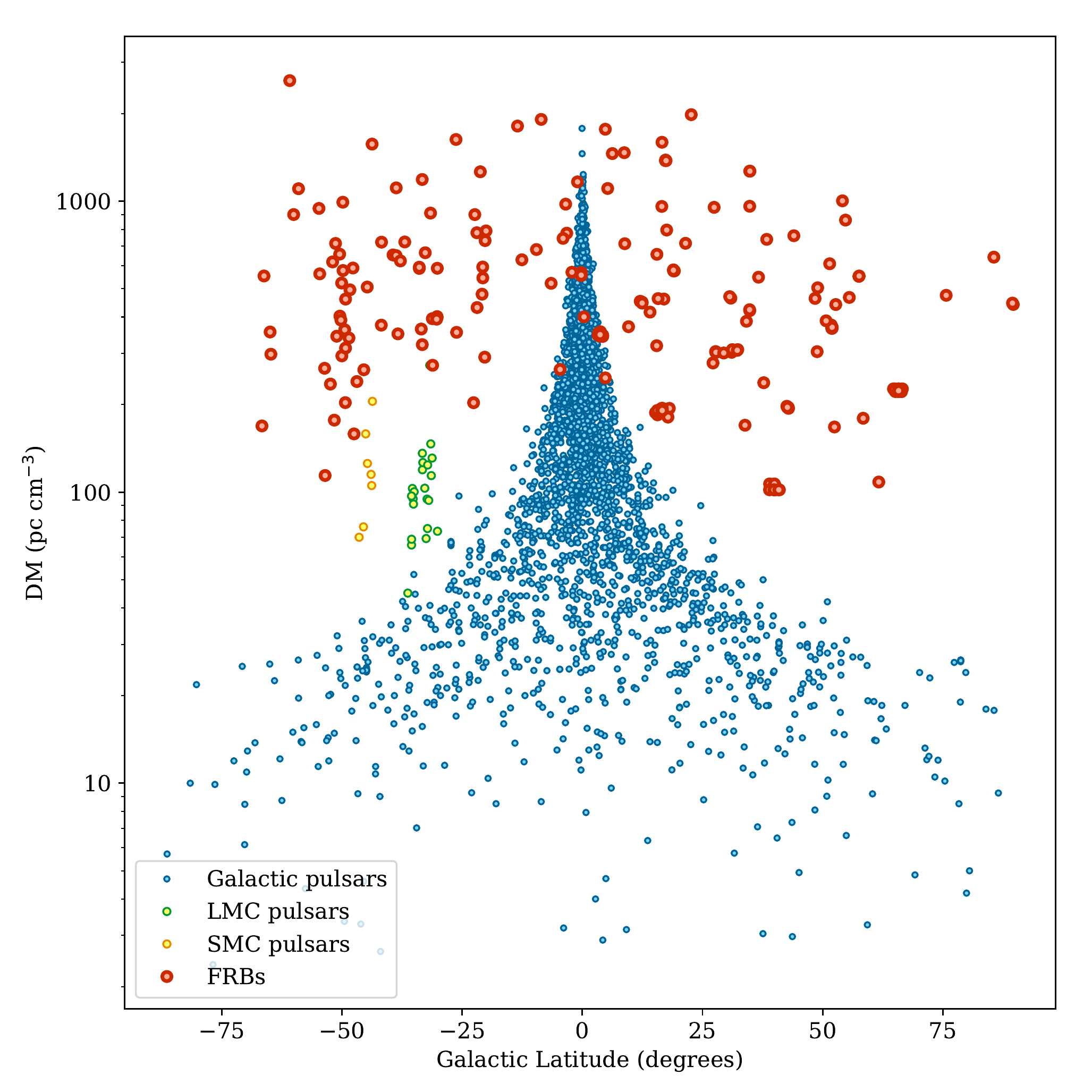}
    \caption{Pulse dispersion measure distinguishes FRBs from pulsars. In this updated version of a plot presented by \citet{cc19}, DM is plotted against Galactic latitude for Galactic pulsars (2751 objects) and pulsars in the LMC (22 objects) and SMC (7 objects), as obtained from the Pulsar Catalogue \citep{psrcat}, as well as for FRBs (284 bursts, from the Transient Name Server service). The DM envelope of the Milky Way is clearly apparent.
    }
    \label{fig:frbdm}
\end{figure}

Examples of FRB dynamic spectra and dispersion-corrected intensity profiles are shown in Figure~\ref{fig:ds}. As of December 2020, there are 284 FRB events (including repeat bursts) recorded by the Transient Name Server (TNS\footnote{TNS: \url{https://wis-tns.weizmann.ac.il/}}), the official IAU repository. Collectively, their DMs show a clear separation from the Galactic pulsar population (Figure~\ref{fig:frbdm}), emphasizing their extragalactic nature. Estimates for the rates of FRB events depend on the sensitivity threshold, with recent estimates ranging from $37\pm8$ events~sky$^{-1}$~day$^{-1}$ brighter than 29~Jy-ms \citep{smb+18} to  $1.7^{+1.5}_{-0.9}\times 10^3$ events~sky$^{-1}$~day$^{-1}$ brighter than 2~Jy-ms \citep{2018MNRAS.473..116K,2018mnras.475.1427b}. However, these estimates are complicated by the fact that most surveys have varying RFI environments, and single-dish radio telescopes have large low-sensitivity sidelobes which could pick up very bright bursts. In any case, the current numbers and rate estimates will soon be rendered obsolete by upcoming results from CHIME, which surveys the sky on a daily transit basis, as discussed further below.

Likewise, understanding the fluence distribution of detected FRBs is complicated for all cases that lack a precise localization from interferometric detection, because of the beam response function (and sidelobes) of single dish telescopes. However, it is safe to say that the current statistics of FRBs are incomplete in every parameter --- fluence, burst width, event rate, DM distribution, repetition, polarization, and more \citep[e.g.,][]{ravi19a}. There is more to be learned.

\section{Repeating Fast Radio Bursts}

\begin{figure}
	\includegraphics[width=0.3\textwidth]{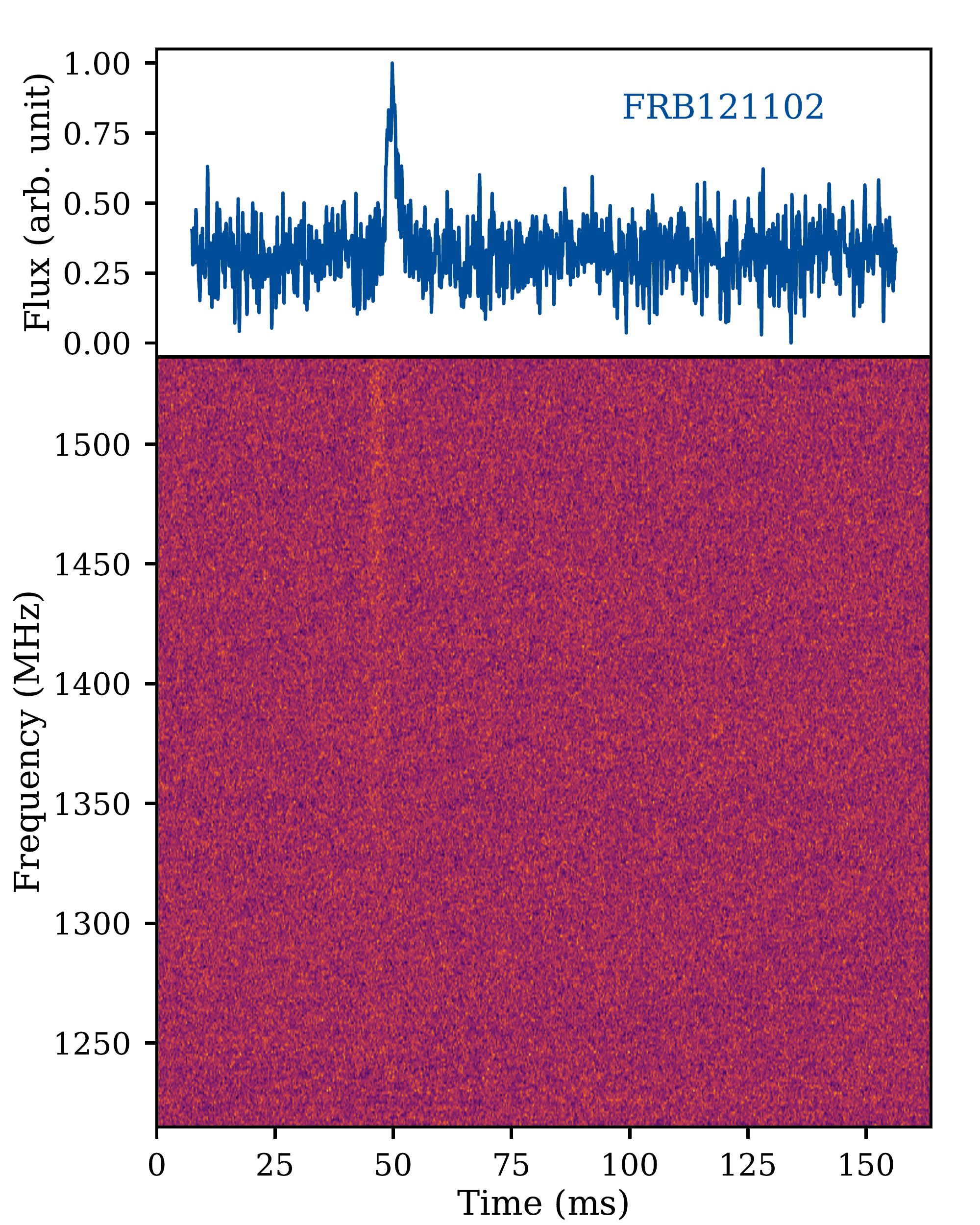}
	\includegraphics[width=0.3\textwidth]{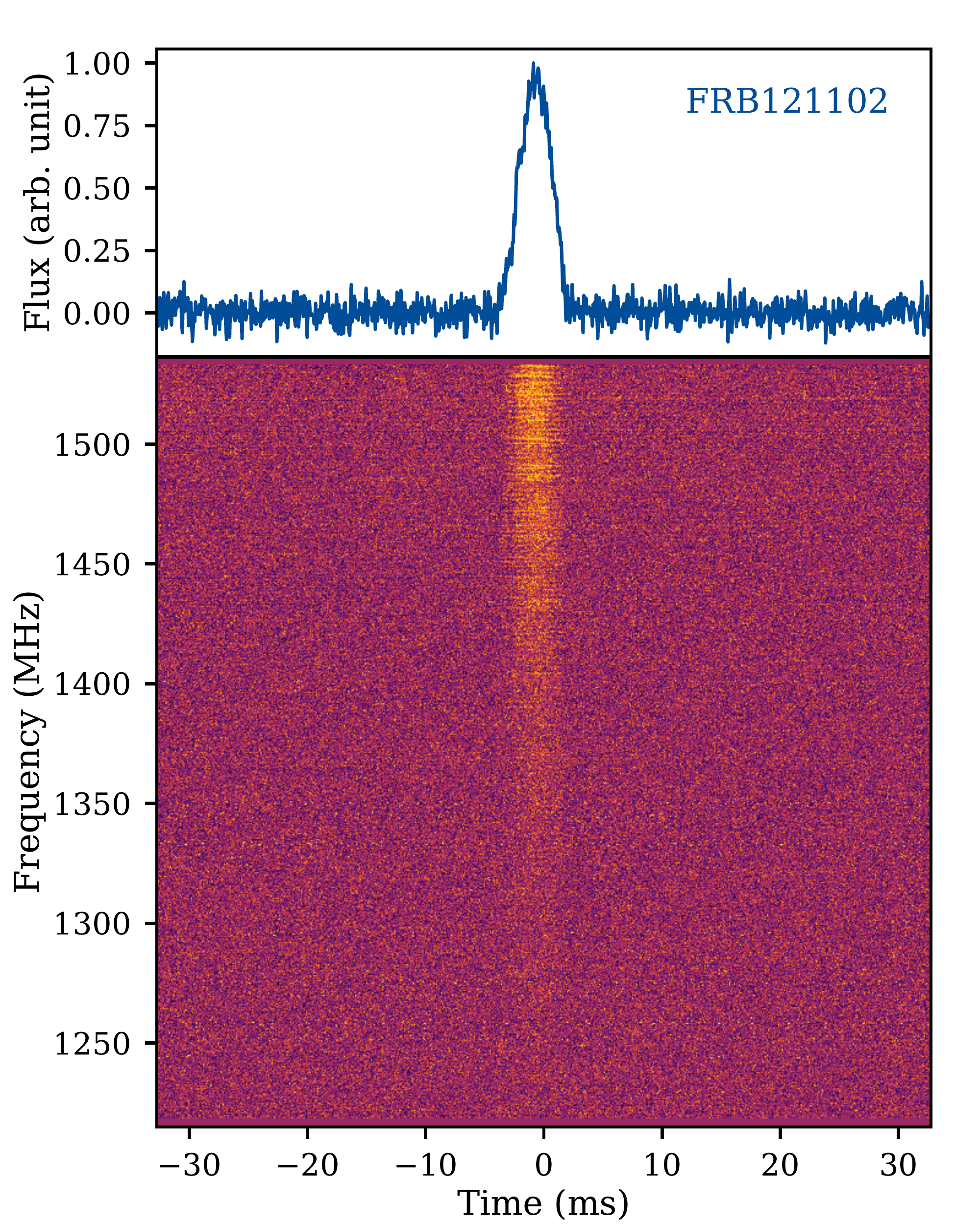}
	\includegraphics[width=0.3\textwidth]{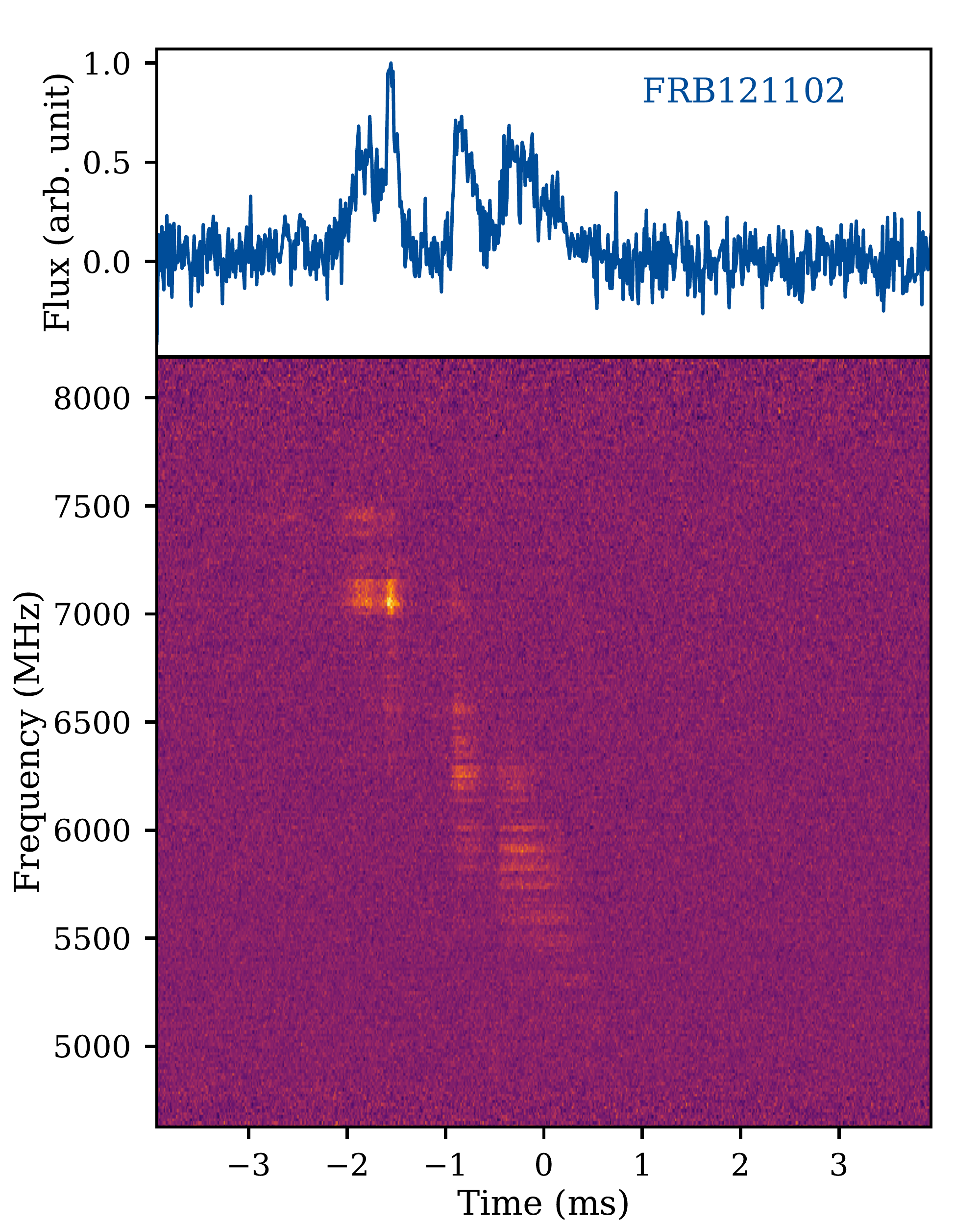}
    \caption{Three bursts from the repeating \rone. As before, lower panels show the dynamic spectrum after removal of the best-fit dispersion sweep, and the upper panels show the frequency-averaged burst profile.
    Left: The faint original detection at Arecibo \citep{sch+14}, showing an inverted spectrum.
    Middle: One of the brightest repeat bursts from \rone\ detected at Arecibo \citep{ssh+16a}.
    Right: A detection at the Green Bank telescope at 4--8~GHz \citep{gsp+18}, showing extensive downward-drifting structure in the time-frequency plane (the ``sad trombone'') after removal of the dispersion sweep.
    }
    \label{fig:rone}
\end{figure}

The discovery that FRB~20121102A (hereafter \rone) was a repeating source (\citealt{ssh+16a,ssh+16b}, and Figure~\ref{fig:rone}) dramatically altered the FRB landscape. While immediately ruling out cataclysmic events as the central engine for at least some FRBs, it suggested the existence of at least two classes of FRBs, in analogy with long and short gamma ray bursts \citep[GRBs,][]{kmf+93}, as well as the possibility that \rone\ was a singular outlier.

A repeating source also allows targeted follow-up observations, and coherent dedispersion at the known DM reveals spectral details at high resolution. As shown in Figure~\ref{fig:rone}, \rone\ shows a diversity of burst morphologies, including single bright sharply-peaked bursts, faint smudges at the detection threshold \citep{gms+19}, and downward-drifting islands of emission in the dynamic spectrum after removal of dispersion \citep{hss+19}, now commonly referred to as the ``sad trombone'' effect and possibly a signature of plasma lensing. \rone\ has been detected at frequencies as high as 4--8~GHz \citep{gsp+18} and as low as 0.4--0.8~GHz \citep{jcf+19}, with event rates as high as 45 bursts in 30 minutes \citep{zgf+18} but no detection of periodicity. Simultaneous multiwavelength observations have revealed no burst counterparts in X-rays, gamma-rays, or optical bands \citep{ssh+16b,hds+17,aaa+18}, although given the millisecond durations of the bursts, the limits are not energetically constraining.

The bursts from \rone\ also appear almost completely linearly polarized, with a Faraday rotation measure that is high ($>10^5$~rad~m${-2}$) and varying over time \citep{msh+18}. The changing RM, without correspondingly significant changes in DM, implies that the projected component of the magnetic field is changing over time, and that the source is embedded in an extreme magneto-ionic medium. Indeed, the only comparable phenomenology, among all the known pulsars in our Galaxy, is seen for the Galactic center magnetar \gcmag\ \citep{2018apj...852l..12d}, in the neighborhood of \sgra.

The discovery of multiple repeating FRB sources at CHIME \citep{chime19repeater, chime19reps,fab+20} confirmed that a fraction of FRBs were detectable as repeating sources, indistinguishable in their DM distributions from the rest of the population, although their burst widths were somewhat larger on average and showed the ``sad trombone'' effect in several cases.

The CHIME repeating FRB~20180916B (hereafter \rthree) is of particular interest. While no repeating FRB source has yet been demonstrated to be periodic, \rthree\ shows periodic windows of detectability \citep{chime20periodic}, with a $\sim$5.2-day window every 16.3~days during which bursts may (or may not) be detected. Such behavior immediately suggests a binary orbit \citep[e.g.,][]{lbg20,iz20}, or precession \citep[e.g.,][]{lbb20,zl20,sobyanin20}, or possibly a very slow underlying rotation period \citep{bwm20} for the central engine. A similar but wider ($\sim$56\%) periodic detectability window has also been claimed for \rone, with a period of $\sim$157 days \citep{rms+20,css+20}.

Recent detections of the repeating FRB~180301 with the Five-hundred-meter Aperture Spherical radio Telescope (FAST) have revealed a high degree of linear polarization, comparable to that seen for \rone. However, these bursts show a wide variety of polarization angle swings \citep{lwm+20}, a puzzling observation that appears inconsistent with models for FRB production in relativistic shocks \citep{lyubarsky14,mm18,beloborodov20}. Instead, \citet{lwm+20} argue that the bursts are produced in neutron star magnetospheres, as suggested by many other models \citep[e.g.][]{katz14,cw16,klb17,zhang17}.

Could all FRBs be repeating sources, or are they two separate classes? It has been apparent from the first FRB rate estimates that cataclysmic phenomena alone cannot account for the event rate \citep{ravi19b}, but in spite of very deep follow-up observations \citep[e.g.,][]{pjk+15,smb+18}, demonstrating that a source never repeats is impossible, and current constraints are limited in nature (\citealt{csrf19,jof+20}; and see \S2.2, ``Follow-up Observations: Trials and Tribulations'' in \citealt{cc19}). \rone, a prolific repeater at higher frequencies \citep{gsp+18}, has been detected only once so far at 400-800~MHz, in spite of daily transits at CHIME \citep{jcf+19}.  FRB~20190711A, an ASKAP-detected burst, has been seen to repeat in deep observations at Parkes, but in an extremely band-limited manner, occupying only 65~MHz of a 3.3~GHz bandwidth \citep{ksf+21}. These observational selection effects argue against a definitive answer to the question at present, although repeating FRBs have not (yet) shown any circularly polarized radio emission, and \cite{dlw+20} propose that feature as a key discriminant between repeating and non-repeating classes, indicative of different radiation mechanisms. However, given the inventiveness of Nature, fast radio bursts might all still have the same underlying engine, or may turn out to be a generic feature from many different source classes.

\section{Localization and Host Galaxies}

The measurement of source distances is a fundamental problem in astronomy, and FRBs have been no exception. Understanding the energetics of FRB central engines, or using FRBs as cosmological probes, requires a distance measurement, most plausibly through the redshift of a host galaxy. However, until large scale surveys started with ASKAP and CHIME, most FRBs were detected by single dish radio telescopes with few-arcminute resolution, while a robust host galaxy association requires few-arcsecond localization \citep{eb17,ebwb18}. Also, in spite of enormous efforts devoted to real-time detection of FRBs and multiwavelength follow-up \citep[e.g.,][]{pbb+15,pbk+17}, no reliable counterparts were identified at other bands.

\begin{figure}
	\includegraphics[width=0.96\textwidth]{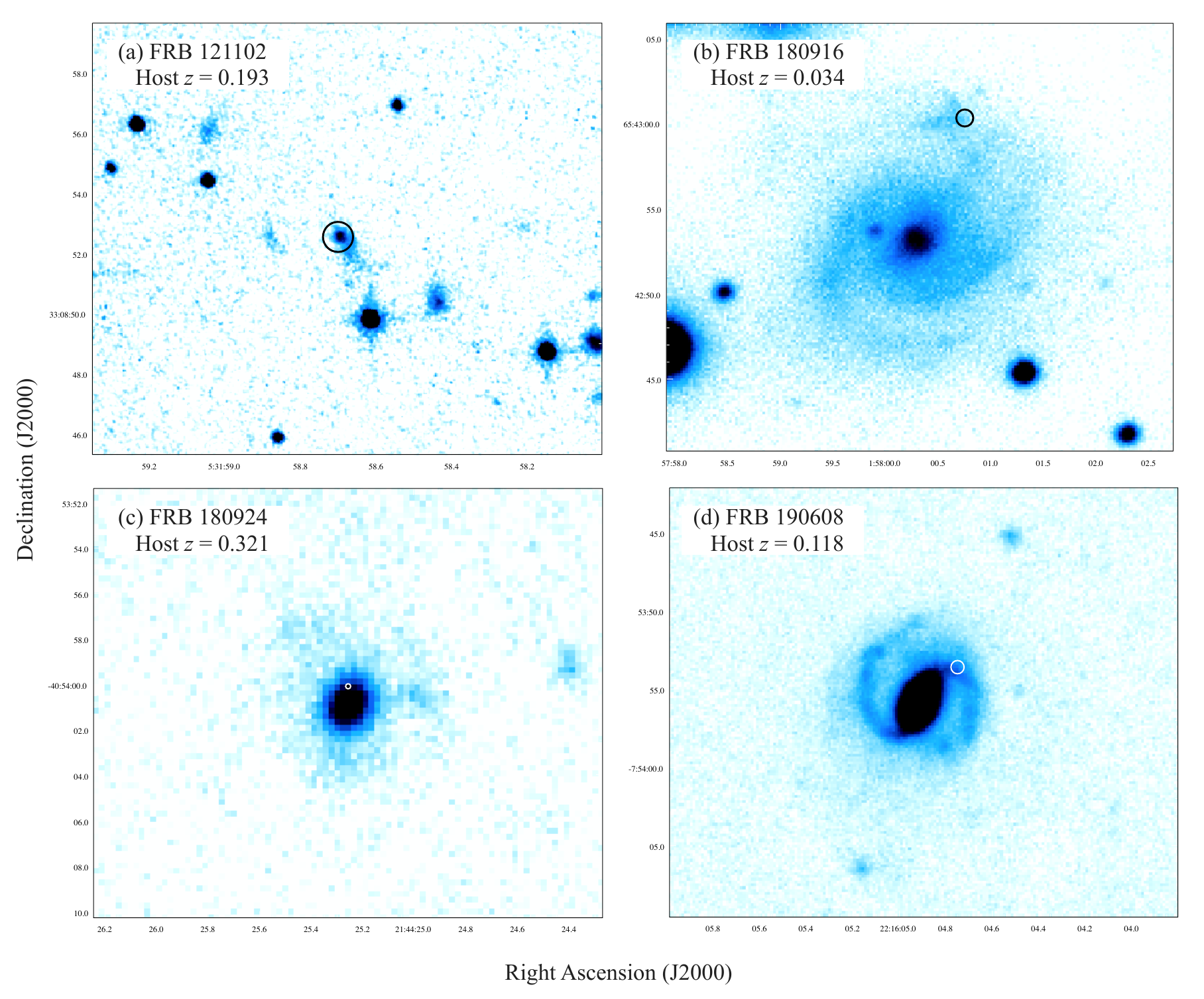}
    \caption{FRB host galaxies show a diversity of morphologies.
    (a) The dwarf host galaxy of the repeating \rone, at $z=0.193$ \citep{tbc+17}, as imaged with the Hubble Space Telescope WFC3 in the F110W (J-band) filter \citep{bta+17}. The localization region is indicated with a circle of radius 0\farcs5; the position is known to milliarcsecond precision \citep{clw+17,mph+17}.
    (b) The nearby spiral host galaxy of the repeating \rthree, at $z=0.034$ \citep{mnh+20}, as imaged with Gemini-North at $r^\prime$-band. As in (a), the localization region is indicated with a circle of radius 0\farcs5, although the position is known to milliarcsecond precision.
    (c) FRB~180924, a single burst detected at ASKAP, is localized to a luminous lenticular or early spiral galaxy at $z=0.321$ \citep{bdp+19}, as imaged with VLT/FORS2 at $g$-band. The localization uncertainty radius is 0\farcs11.
    (d) FRB~190608 is another single ASKAP-detected burst, localized to a nearby spiral galaxy at $z=0.118$ \citep{mpm+20} with a localization uncertainty radius of 0\farcs42, as imaged with VLT/X-shooter at $g$-band.
    }
    \label{fig:hostgalaxies}
\end{figure}

The first direct localization of a fast radio burst \citep[\rone,][]{clw+17} relied on VLA interferometric follow-up of the first source known to repeat. The identification of a star-forming dwarf galaxy host (Figure~\ref{fig:hostgalaxies}) and the measurement of its redshift, $z=0.193$ \citep{tbc+17} allowed a firm estimate of the burst energetics from the fluence $A_{\nu}$ of detected bursts,
\[
E_{\rm burst} = 4\pi D^2 (\delta\Omega/4\pi) A_{\nu} \Delta\nu
\approx 10^{38}\, {\rm erg}\,(\delta\Omega/4\pi) D_{\rm Gpc}^2  (A_{\nu} / 0.1\ {\rm Jy\ ms}) \Delta\nu_{\rm GHz},
\]
where the distance $D \sim 1$~Gpc and the unknown beaming solid angle $\delta\Omega$ scales the total energy requirement. Note, however, that while a narrower beam reduces the energy budget per source, it requires a corresponding increase in the number of such sources. In any case, the energetics are compatible with neutron stars or other compact object progenitors. In addition, a persistent radio source is associated with the bursts at milliarcsecond scales \citep{clw+17,mph+17}. While the persistent radio source is not inconsistent with a (weak) active galactic nucleus \citep{vrb+17}, it is naturally explained as a nebula surrounding a very young magnetar in a concordance model for \rone\ \citep{mm18}, consistent with its association with a star-forming region in the host galaxy \citep{bta+17}.

However, this self-consistent picture has not proven to be universal. Interferometric localizations of one-off FRBs with ASKAP \citep{bdp+19,pmm+19,mpm+20,bsp+20,hps+20} as well as the DSA-10 \citep{rcd+19} and the VLA \citep{lbp+20} led to the identification of a diversity of host galaxies in morphology and metallicity, as well as diverse environments within the galaxies themselves, as illustrated for a sample in Figure~\ref{fig:hostgalaxies}.

Even the second repeating FRB to be localized, the CHIME-detected \rthree, was localized with EVN observations to a nearby, massive spiral galaxy \citep[][see Figure~\ref{fig:hostgalaxies}]{mnh+20}, with no associated persistent radio source detected in spite of its low redshift ($z=0.034$) and distance (149~Mpc). With high-resolution optical observations, \citet{2020arXiv201103257T} show that the source, with its periodic detectability windows described above, is not associated with a star-forming region and inconsistent with even a runaway magnetar, and suggest that the source may instead be a high-mass X-ray (or gamma-ray) binary.

\begin{figure}
	\includegraphics[width=0.48\textwidth]{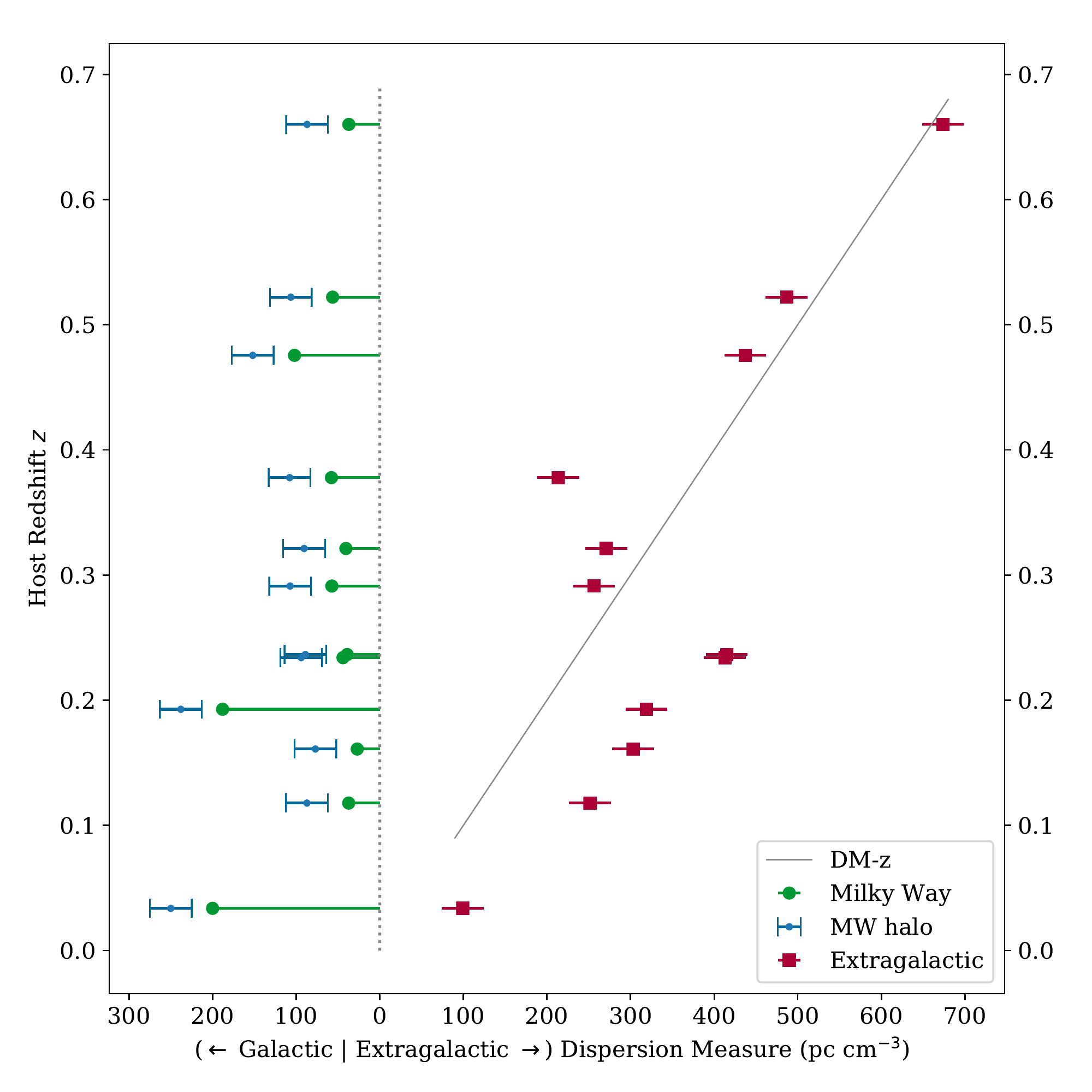}
    \caption{The Galactic and extragalactic components of FRB dispersion measure plotted against the FRB host galaxy redshift. The Milky Way contribution from NE2001 is shown on the left, along with a nominal Milky Way halo contribution of $50 \pm 25~\DMunits$. The extragalactic \DM\ includes an unknown host galaxy contribution, with the uncertainty due to the Milky Way halo contribution shown for emphasis. A line is plotted to indicate an approximate relationship DM $\approx 1000\;z\;\DMunits$ for the IGM at low $z$ \citep[e.g.,][]{mcq14}, to which host galaxy contributions should add on.
    }
    \label{fig:dmz}
\end{figure}

Alongside the quest to understand the central engine behind FRB emission, the promise of FRBs with well-measured distances is the ability to directly probe the baryon content of the IGM \citep[e.g.,][]{iok03,ino04,mcq14,pz19}. At low redshift, the vast majority of the baryon content of the universe is not seen \citep[the so-called ``missing baryon problem''; e.g.,][]{bregman07} but is believed to lie in gaseous filaments between galaxy clusters. As shown in Figure~\ref{fig:dmz}, the extragalactic DM component of FRBs correlates with the host galaxy redshift.

Note, however, the scatter about a simple linear trend in Figure~\ref{fig:dmz}. The measured DM of each FRB includes contributions from the Milky Way, subject to modeling uncertainties in NE2001 \citep{cl02} or YMW16 \citep{ymw17}. The contribution of the Milky Way halo is not well known, with a range of plausible values spanning $50 \pm 25 \DMunits$ \citep[e.g.,][]{pz19}. The host galaxy (and its halo) will also contribute similar amounts to the total DM, and these contributions (weighted for redshift) have to be accounted for before a redshift-DM relationship can be measured. In a first such effort, \citet{mpm+20} use a carefully-selected sample of FRBs to estimate the cosmic baryon density and show consistency with cosmological predictions. The uncertainties are large at present, but can be refined with enough well-measured FRBs. In addition, FRBs also probe galaxy or cluster halos along the line of sight \citep[e.g.,][]{cvo+20} and a large enough sample of FRBs may allow detailed tomographic modeling of nearby clusters \citep{rbb+19}. In order to realize the promise of FRBs as extragalactic probes, it is crucial to improve modeling of the electron density distribution in the Galaxy \citep[e.g.][]{occ20} and the Galactic halo.

\section{A Galactic FRB and the FRB central engine}

As discussed above, there are a diverse range of models for FRBs \citep[see, e.g.,][]{z20} but compact objects (and specifically young magnetars) are implicated in most viable ones. The association of a compact persistent radio source with \rone, along with its high rotation measure, favors models where the FRB emission arises from relativistic shocks in the nebula surrounding the neutron star, but such a persistent source is not seen for the nearby repeating \rthree. In contrast, the periodic detectability of \rthree\ (and possibly \rone) can be naturally explained by an orbit, precession, or very slow rotation; and the varying polarization angle swings for the repeating FRB~180301 suggest a magnetospheric origin for the bursts. The situation appears (delightfully) unsettled, although the FRB sample with secure host galaxy associations is consistent with an origin in a population of magnetars \citep{brd20,hps+20}. That naturally raises the prospect of detecting FRBs from magnetars in our own Galactic backyard, although with the caveat that such a burst might be so bright that it would be flagged and excised as radio interference.

\begin{figure}
	\includegraphics[width=0.48\textwidth]{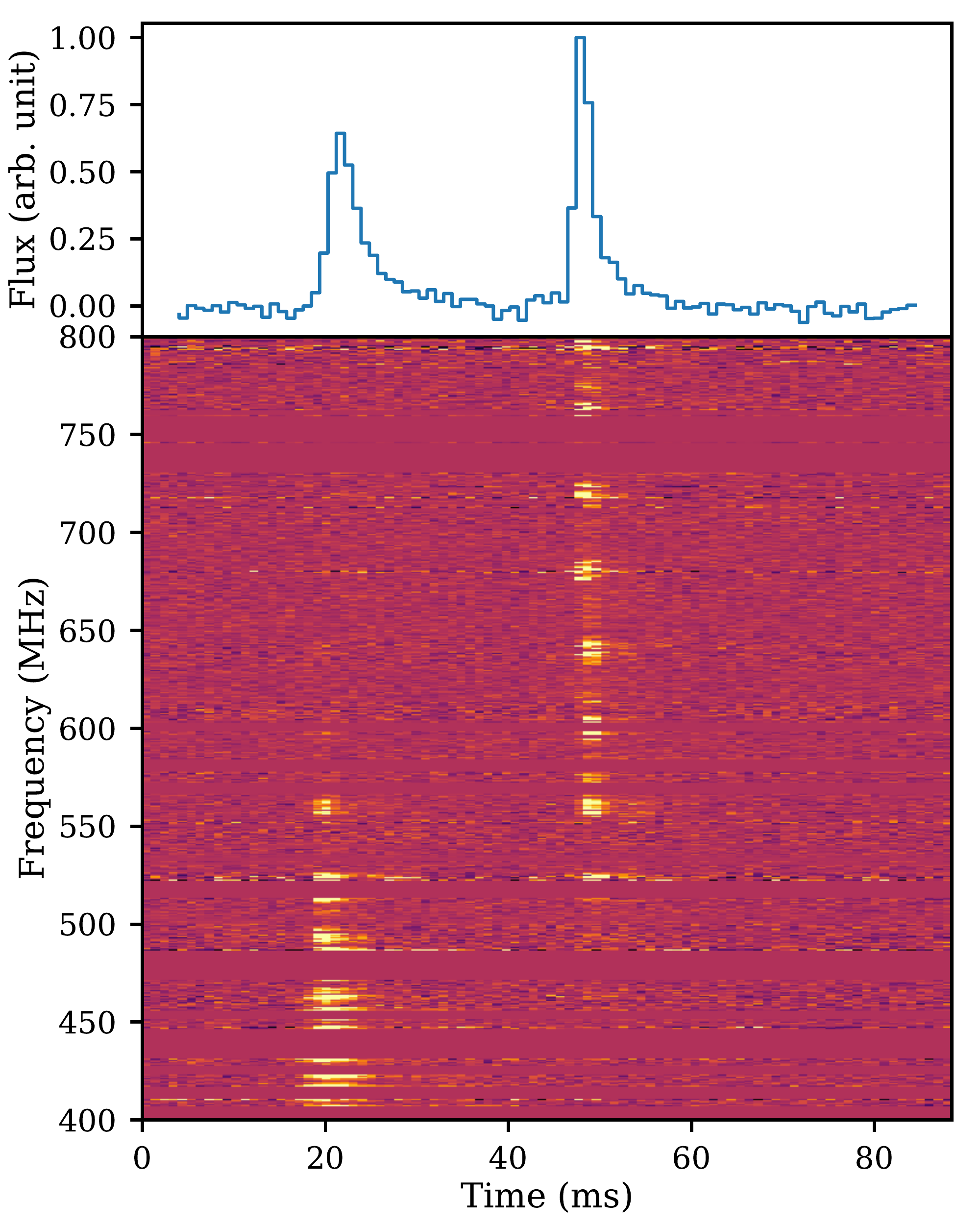}
	\includegraphics[width=0.48\textwidth]{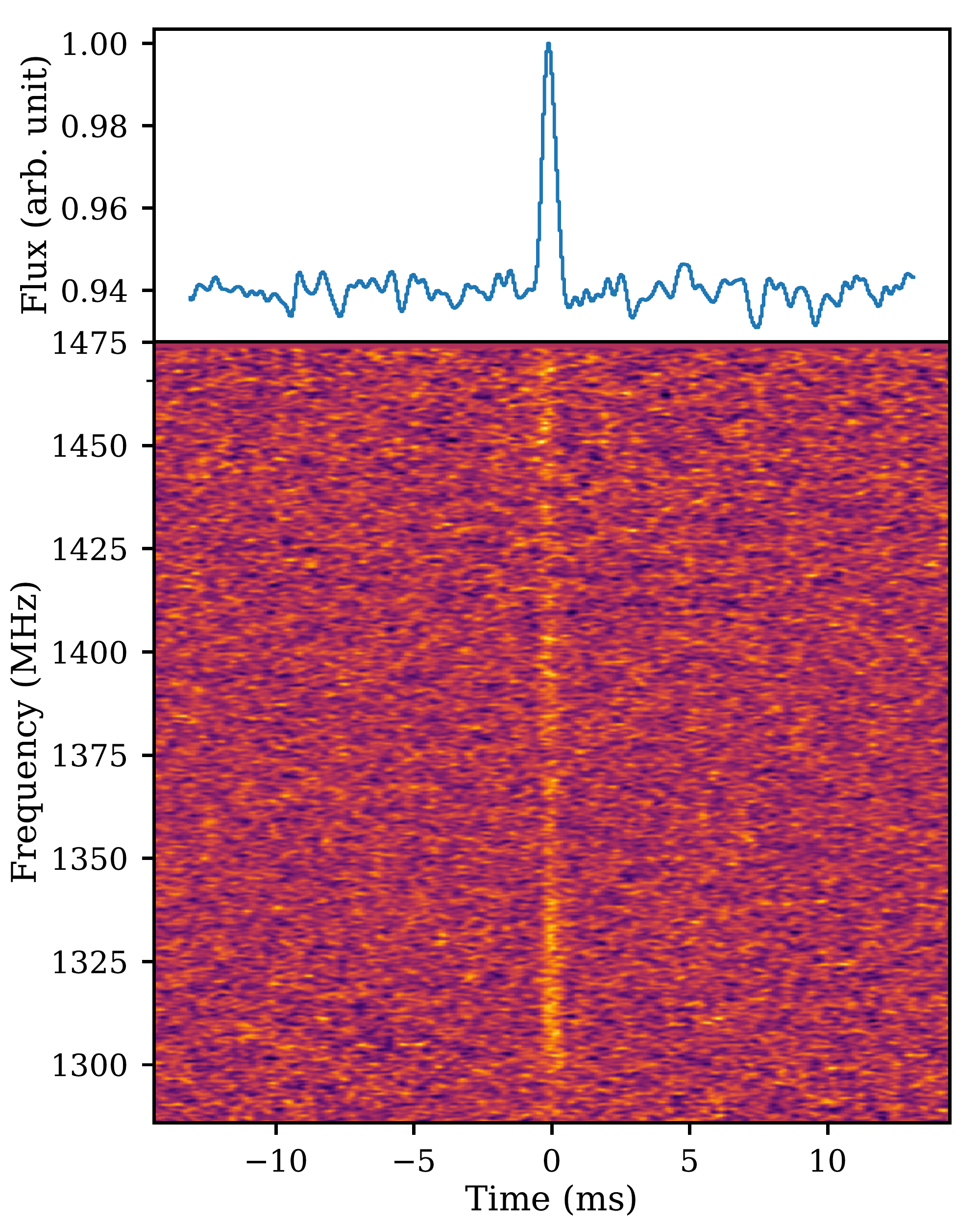}
    \caption{Radio burst from \sgr, as detected at (left) CHIME at 400--800 MHz \citep{chime20magnetar}, and (right) STARE2 at 1.4~GHz \citep{brb+20}. Time axis labels have an arbitrary offset; the higher (second) peak of the CHIME burst is aligned with the STARE2 burst in time after correcting for dispersive delays at the different frequencies.
    }
    \label{fig:sgrflare}
\end{figure}

Such a prospect has been spectacularly confirmed with the recent detection of radio bursts from the Galactic magnetar \sgr. Magnetars are known to produce gamma-ray bursts and giant X-ray flares, quasi-periodic activity, periodic radio pulses, and more \citep{kb17}, primarily driven by the potential energy stored in their extreme magnetic fields (B$>10^{14}$~G). While the magnetar \sgr\ was in an extended ``active'' phase, producing hundreds of X-ray flares (see \citealt{ww20} for a timeline of events), an extremely intense two-component radio burst was detected by CHIME at 400-800~MHz \citep{chime20magnetar} with a combined fluence of 700~kJy~ms and a DM of 332.7~$\DMunits$. Coincident with that burst, a burst was detected at the STARE2 array at 1.4~GHz \citep{brb+20} with a fluence of 1.5~MJy~ms and the same DM. The burst dynamic spectra are shown in Figure~\ref{fig:sgrflare} after removal of the pulse dispersion; these bursts were bright enough to have been detected and unambiguously classified as FRBs if they had been beamed towards us from nearby galaxies.

Several satellites also detected a bright, hard X-ray burst coincident with the radio burst (after accounting for pulse dispersion delays) among a forest of other X-ray bursts \citep[for example, AGILE;][]{tcu+20},  suggesting that the radio emission had a much narrower beaming angle than the X-rays did. Observations at FAST also placed upper limits on any pulsed radio emission \citep{lzw+20}, although a few fainter bursts were detected \citep{ksj+20}, suggesting a possible continuum of bursts extending down to much weaker fluences (at least 7 orders of magnitude).

While the detection of radio bursts from \sgr\ confirms that magnetars produce at least some fraction of FRBs, there still exists a 30$\times$ difference in intrinsic energy between the bright burst seen at STARE2 and even the weaker known FRBs, so the provenance of the brightest FRBs remains a somewhat open question. \citet{mbsm20} argue that the properties of the coincident radio and X-ray flares from \sgr\ are consistent with the predictions of the synchrotron maser shock model rather than arising in the magnetosphere, but propose an additional population of extremely active magnetars with even stronger magnetic fields to account for the observed FRB rate and repeating fraction.  A unified model for the weaker Galactic magnetar bursts and the brighter cosmological FRBs is proposed by \citet{lkz20}, who invoke magnetic disturbances propagating outwards from near the magnetar surface to produce radio bursts.

We note, for now, that the number of models exceeds the number of observational constraints and that no single model yet explains all the observed phenomenology of FRBs, including the volumetric rates, morphology differences, repeatability, periodic detectability windows, polarization properties, and more. The prospect remains open that FRB emission could arise from multiple engines, or from several closely related source classes, although the likelihood is high that magnetars (or other neutron stars) are central to the process.

\section{Future Prospects}

In less than a decade, fast radio bursts have gone from a single debated curiosity to a diverse extragalactic population with established host galaxies and energy scales. While a wide range of models remain viable, the central engines of FRBs are likely to involve energetic young magnetars, as confirmed by the recent discovery of a Galactic analog to these extragalactic bursts. However, we caution against a rush to judgement on the mechanisms and classification of all FRBs: at this point, it remains plausible that there could be one dominant central engine, as well as the possibility that radio bursts are a generic feature produced by many different mechanisms.

FRBs are already being used to probe the halos of other galaxies \citep{pz19,cvo+20} and to address the distribution of baryons in the IGM \citep{mpm+20}. In future, the availability of a large sample of well-characterized FRBs with host galaxies may be used to address a broad range of topics \citep[e.g.,][]{rbb+19}, including measurement of the IGM magnetic field \citep{arg16}, the cosmological baryon density at intermediate redshifts \citep{wwg+18}, and the era of helium reionization \citep{cfs19}. FRB microlensing can reveal the presence of massive compact halo objects as a constituent of dark matter \citep{mkdk16}.

In order to live up to this promise, many more FRBs need to be detected. In this regard, the ongoing transit sky survey with CHIME is likely to be transformative, with hundreds of detected FRBs and a well-defined sky exposure map that will allow for robust statistical analyses. Further, while FRB detections are interesting in their own right, the bursts need to be localized to $\sim$arcsecond precision and host galaxy redshifts measured in order to unlock their potential as probes of the IGM density and magnetic field. Interferometric surveys with ASKAP, the DSA-10, and the VLA have already demonstrated localizations from one-off detections, with similar efforts underway at other interferometers (UTMOST-2D, DSA-110, MeerKAT, and more). In future, very long baseline interferometry at the EVN, VLBA, and eventually the CHIME outriggers have the potential to associate one-off FRB detections not just with host galaxies, but with specific regions that allow constraints on their progenitors, as already demonstrated for repeating FRBs.  Closer to home, we also emphasize the importance of improving the modeling of our Galaxy and Galactic halo, which otherwise impose systematic errors on every FRB line of sight. The future of science with fast radio bursts appears bright.

\section*{Acknowledgements}

The author thanks Jim Cordes, Emily Petroff, and many other colleagues, especially in the CHIME and NANOGrav collaborations, for extensive discussions and access to data.
The author acknowledges support from the National Science Foundation (AAG 1815242), and is a member of the NANOGrav Physics Frontiers Center, which is supported by the National Science Foundation award number 1430284.


\input{msbbl.txt}

\bsp	
\label{lastpage}
\end{document}